\documentclass{elsart}
\usepackage{graphicx}
\usepackage{bm}
\usepackage{amsmath}
\usepackage[centerlast]{subfigure}
\begin{document}
\begin{frontmatter}
\title{Nonlinearly driven transverse synchronization in coupled
chaotic systems}

\author{Massimo Cencini
$^{(a)}$
and Alessandro Torcini $^{(b)}$}
\address{$^{(a)}$ ISC-CNR Via dei Taurini 19, I-00185 Roma, Italy and
SMC-INFM, Dip. Fisica, Universit\`a di Roma ``La Sapienza'', p.zzle Aldo Moro
2, I-00185 Roma, Italy} \address{$^{(b)}$
ISC-CNR, Sezione Territoriale di Firenze, and
Istituto Nazionale di Ottica Applicata, L.go E. Fermi, 6 -
I-50125 Firenze, Italy}
\vskip 1.5cm
\begin{abstract}
\noindent
Synchronization transitions are investigated in coupled chaotic maps.
Depending on the relative weight of linear versus nonlinear
instability mechanisms associated to the single map two different
scenarios for the transition may occur. When only two maps are
considered we always find that the critical coupling $\varepsilon_l$
for chaotic synchronization can be predicted within a linear analysis
by the vanishing of the transverse Lyapunov exponent $\lambda_T$.
However, major differences between transitions driven by linear or
nonlinear mechanisms are revealed by the dynamics of the transient
toward the synchronized state.  As a representative example of
extended systems a one dimensional lattice of chaotic maps with
power-law coupling is considered. In this high dimensional model
finite amplitude instabilities may have a dramatic effect on the
transition.  For strong nonlinearities an exponential divergence of
the synchronization times with the chain length can be observed above
$\varepsilon_l$, notwithstanding the transverse dynamics is stable
against infinitesimal perturbations at any instant. Therefore, the
transition takes place at a coupling $\varepsilon_{nl}$ definitely
larger than $\varepsilon_l$ and its origin is intrinsically
nonlinear. The linearly driven transitions are continuous and can be
described in terms of mean field results for non-equilibrium phase
transitions with long range interactions. While the transitions
dominated by nonlinear mechanisms appear to be discontinuous. \\
\noindent
\end{abstract}

\end{frontmatter}

\section{Introduction}
\label{sec:1}
Synchronization is a phenomenon observed in many different contexts,
ranging from epidemics spreading \cite{epidemie} to neuro-sciences
\cite{neuro}. The analysis of simple models pursued in the last
two decades has clarified several aspects of
synchronization (for a comprehensive review on the subject see
Ref.~\cite{PikoBook}).

Particularly interesting both from a theoretical and an applied point
of view is synchronization of chaotic systems.  This is a
phenomenon known since many years~\cite{first} with important
applications, e.g., in secure communications of digital signals, laser
dynamics etc. A simple but not trivial framework to study chaotic
synchronization is represented by discrete-time dynamical
systems. Within this context, the essence of the phenomenon can be
captured already by considering two coupled identical maps:
\begin{eqnarray}
x^{t+1} & = & (1-\varepsilon) f(x^t) + \varepsilon f(y^t) \nonumber\\
y^{t+1} & = & \varepsilon f(x^t) + (1-\varepsilon) f(y^t) 
\label{eq:2map_1}\,,
\end{eqnarray}
where $f(x)$ is a chaotic map of the unit interval and $\varepsilon$
the coupling constant.  For coupling larger than a critical value
$\varepsilon_l$ a transition from a desynchronized state to a
completely synchronized one along a chaotic orbit $x^t$ is observed,
i.e. $x^t \equiv y^t$ for $\varepsilon > \varepsilon_l$.
The coupling $\varepsilon_l$ can be predicted
within a linear analysis as the value of $\varepsilon$ for which the
Transverse Lyapunov Exponent (TLE) $\lambda_T$, ruling the instabilities
transversal to the line $x=y$, vanishes. For model (\ref{eq:2map_1})
$\lambda_T=\ln(1-2\varepsilon_l)+\lambda_0$ \cite{PikoBook}, and consequently 
\begin{equation}
\varepsilon_l = \frac{1-{\rm e}^{-\lambda_0}}{2} \,,
\label{eq:epsc}
\end{equation}
being $\lambda_0$ the Lyapunov Exponent (LE) of the single (uncoupled) map.  

While in low dimensional systems the basic mechanisms ruling synchronization,
in the absence of strong nonlinear effects, have been well established 
since long time~\cite{PikoBook,KP89,PG91}, the studies devoted to
synchronization in extended systems are still limited to few 
specific examples~\cite{J95,baroni,AP02,anteneodo}.
In the latter case, two main frameworks have been considered.

On one hand, the investigation of mutual synchronization between two
replicas of a spatially extended system has been pursued for
systems coupled either via local interaction \cite{AP02} or via 
spatio-temporal noise \cite{baroni}.  In both cases,
when the amplitude of the coupling (noise) overcomes a certain
threshold, synchronization is observed: after a transient, the two
systems follow the same spatio-temporal chaotic (stochastic) orbit.

On the other hand, another commonly observed situation corresponds to
self synchronization of elements belonging to the same system.
This occurs for instance in large collections of coupled elements such
as populations of neurons \cite{neurosynch}, Josephson junctions
\cite{jose} or cardiac pacemaker cells \cite{cuore}. In these systems
the interaction among the elements can range from nearest neighbors
to globally coupled. Therefore, a model able to encompass both these
limiting cases is certainly of interest.  A good candidate is
represented by the following coupled map lattice with power-law
interactions~\cite{PV94,lep-torc,anteneodo}
\begin{equation}
x_i^{t+1}= (1-\varepsilon) f(x_i^t) + 
\frac{\varepsilon}{\eta(\alpha)} \sum_{k=1}^{L^\prime}
\frac{f(x_{i-k}^t)+f(x_{i+k}^t)}{k^\alpha}\,,
\label{eq:cml}
\end{equation}
which reduces to globally coupled maps (GCM's) \cite{gcm} for
$\alpha = 0$, and to standard coupled map lattices (CML's)
with nearest neighbor coupling \cite{cml} in the limit
$\alpha \to \infty$. In equation (\ref{eq:cml}), $t$ and $i$ are the discrete temporal and
spatial indexes, $L$ is the lattice size ($i=1,\dots,L$), $x_i^t$ the
state variable, $\epsilon \in [0:1]$ measures the strength of the
coupling and $\alpha$ its power-law decay. Since the sum extends up to
$L^\prime = (L-1)/2$ the model is well defined only for odd
$L$-values, and $\eta(\alpha)=2\sum_{k=1}^{L^\prime} k^{-\alpha}$ is a
normalization factor. Periodic boundary conditions are assumed. 

The synchronization between replicas of the same spatial
system has been the subject of a more systematic study than self
synchronization and is now rather well understood~\cite{baroni,A01,AP02,G03,munoz}. 
Two different mechanisms of mutual synchronization have been identified
according to the predominance of linear versus nonlinear effects.  
Nonlinearly dominated transitions are observed 
when the local dynamics is ruled by a discontinuous (e.g the
Bernoulli map) or "almost discontinuous" maps (e.g. maps possessing 
very high values of the first derivatives). In these cases a linear analysis
is no more sufficient to fully characterize the transition,
because the instabilities associated with finite amplitude 
perturbations may desynchronize the system. 
In both cases the transition to the synchronized state
is typically of the second order. However, 
depending on the linear or non-linear
nature of the prevailing mechanism two different universality classes
characterize the transition itself. For continuous maps (e.g. logistic
maps) critical exponents associated with the Multiplicative Noise (MN)
universality class are usually found~\cite{munoz_review}, while for (almost) 
discontinuous maps the transition belongs to the Directed Percolation (DP)
class~\cite{haye}.  

The aim of the present work is to clarify the effect of strong
nonlinearities in the synchronization phenomenon, in particular in
the case of self synchronization.

As a first point, we show that strong nonlinearities may induce
nontrivial effects also in low dimensional systems such as two coupled
maps.  In particular, we find that even though the critical coupling
$\varepsilon_l$ can be always predicted within the linear framework,
the transient dynamics preceding the synchronization can be strongly
affected by finite scale instabilities.  To properly characterize the
latter we introduce a new indicator, the Finite Size Transverse
Lyapunov Exponent (FSTLE), which generalizes the concept of Finite Size
Lyapunov Exponent (FSLE)~\cite{ABCPV96,BCFV02,CT01} to the transverse
dynamics.  The FSTLE extends the definition of TLE to finite
perturbation and is therefore able to discern linearly from
nonlinearly dominated transitions.  Depending on the behavior of the
FSTLE at finite amplitudes two different class of maps are singled
out\: {\it class I} maps characterized by a decreasing FSTLE at any
finite scale, and {\it class II} maps that present a peak in the FSTLE
for some finite amplitude value. Moreover, in the proximity of the
transition the shape of the probability density function (PDF),
$P(\tau)$, of the synchronization times $\tau$ depends on the dynamics
at finite scales and on the multifractal properties of the map.  For
class I maps two cases can be identified. For maps, such as the
symmetric tent map and the logistic map at the crisis, which are
characterized by the vanishing of fluctuations of the finite time LE
at long times, the PDF exhibits a fast falloff at large $\tau$'s. In
particular, for the symmetrical tent map we provide an analytical
expression for $P(\tau)$.  For maps, such as the skew tent map,
exhibiting {\it modulational intermittency} \cite{PikoBook} (which is
related to the persistence of fluctuations of the finite time LE also
in the long time limit), $P(\tau)$ becomes an inverse Gaussian
distribution \cite{Feller} originating by the diffusive motion of the
(logarithm of the) perturbation in transverse space.  For class II
maps the PDF's display an
exponential tail at long times which, differently from the previous
case, is due to nonlinear effects.

As far as spatially extended systems are concerned, Anteneodo et
al.~\cite{anteneodo} have performed a linear stability analysis of
model (\ref{eq:cml}) obtaining analytically the critical coupling
$\varepsilon_l$ for the transition.  This prediction works
perfectly for class I maps, as found for the coupled logistic
maps~\cite{anteneodo}. However, as shown in the present paper, 
linear analysis may fail in class II maps. In particular, for large system
sizes ($L \gg 1$) an exponential divergence of the synchronization
times with $L$ can be observed even for negative TLE, as a result 
in the limit $L\to \infty$ the transition takes place at a critical coupling
$\varepsilon_{nl} > \varepsilon_l$.  Finite scale instabilities are
the key elements for observing this {\it nonlinear synchronization
transition}. Similarly to the case of diffusively coupled map lattices
it is possible to link the synchronization transition of  CML
with power law coupling with non-equilibrium phase transitions.
Indeed as shown in Ref.~\cite{munoz_review} synchronization of 
replicas of CML with short range coupling belong either to
the MN or to the DP universality classes.
Here we shall discuss the connection of self synchronization of
model (\ref{eq:cml}) with long range spreading processes~\cite{haye,haye05}.

The material is organized as follows. Sect.~\ref{sec:2} is devoted to
synchronization of two coupled maps.  In Sect.~\ref{sec:3} coupled
maps with power law coupling are examined.  Finally in
Sec.~\ref{sec:4} we briefly summarize the reported results.

\section{Synchronization of two coupled identical maps}
\label{sec:2}                   
To analyze the synchronization of two coupled maps (\ref{eq:2map_1})
it is useful to introduce the following variables:
$$ u^t = \frac{x^t+y^t}{2} \qquad w^t = \frac{x^t-y^t}{2} \, ,$$
in terms of which Eq.~(\ref{eq:2map_1}) can be rewritten  as
\begin{eqnarray}
u^{t+1} & = & \frac{1}{2} \left[f(u^t+w^t) + f(u^t-w^t)
\right]\nonumber\\
w^{t+1} & = & \left( \frac{1}{2}-\varepsilon \right) 
\left[f(u^t+w^t) - f(u^t-w^t)
\right]
\label{eq:diff}\,.
\end{eqnarray}
It is easily checked that the synchronized solution (corresponding to
$w^t=0$ and $u^t=f(u^t)$) is an admissible solution of (\ref{eq:diff})
for any value of the coupling $\varepsilon$. The stability of such a
solution can be studied by considering the linear dynamics of an
infinitesimal perturbation $\delta w^t$ of the synchronized
state. This evolves according to the linearized equation:
\begin{equation}
\delta w^{t+1}  =  (1-2\varepsilon) f^\prime(u^t) \delta w^t
\label{eq:tang_diff}\,,
\end{equation}
where $f^\prime=df/dx$.
Clearly the stability of the state $w^t=0$ is controlled by the sign
of the transverse (or conditional) Lyapunov exponent
\begin{equation}
\lambda_T=\ln(1-2\varepsilon)+\lambda_0  \,,
\label{eq:lambdat}
\end{equation}
where $\lambda_0$ is the Lyapunov exponent of the single map 
$f(x)$. Therefore, by requiring $\lambda_T=0$ one obtains the critical
coupling $\varepsilon_l$ (\ref{eq:epsc}) above which the synchronized
state is linearly stable.  Notice that for $\varepsilon \ge \varepsilon_l$,
$\lambda_T$ coincides with the second Lyapunov exponent $\lambda_2$ of
the coupled system (\ref{eq:2map_1}).  

The synchronization transition has been mainly studied for 
continuous maps, here denoted as {\it class I} maps, 
in particular for the logistic map
$f(x)=4 x(1-x)$ at the crisis \cite{KP89} and the skew tent map
\cite{PG91}
\begin{equation}
f(x)=\left\{\begin{array}{ll}
  x/a &{\rm if} \quad 0\le x \le a\\
  (1-x)/(1-a) &{\rm if} \quad a < x \le 1
\end{array}
\right. \,.\label{eq:skt}
\end{equation}
It should be noticed that inside this class of maps one has to
distinguish between two situations according to the behavior of the
fluctuations of the finite time Lyapunov exponents in the long time
limit. The vanishing of such fluctuations is observed for the
symmetric tent map and the logistic one at the crisis, while their
persistence characterizes the skew tent map that represents the
generic case \cite{PikoBook}.

Recent studies however pointed out that discontinuous or ``almost''
discontinuous maps (i.e. maps with $|f^\prime|\gg 1$ in some point of
the definition interval), here termed {\it class II\/} maps, may give
rise to non trivial interesting phenomena. For instance, the
synchronization transition for two replicas of class II CMLs is not
driven, as usual, by infinitesimal perturbations, but by the
instability (spreading) of finite amplitude perturbations
\cite{baroni,AP02}.  This nonlinear synchronization is
linked to the so-called {\it stable chaos}~\cite{PLOK93}: a
phenomenon characterized by information propagation even in the
absence of chaos~\cite{TGP95,CT01}.  Moreover, as we shall show in the
following also the self synchronization of a single CML can be
strongly affected by nonlinearities.
\begin{figure}[t!]
\centering
\includegraphics[draft=false, scale=.42,clip=true]{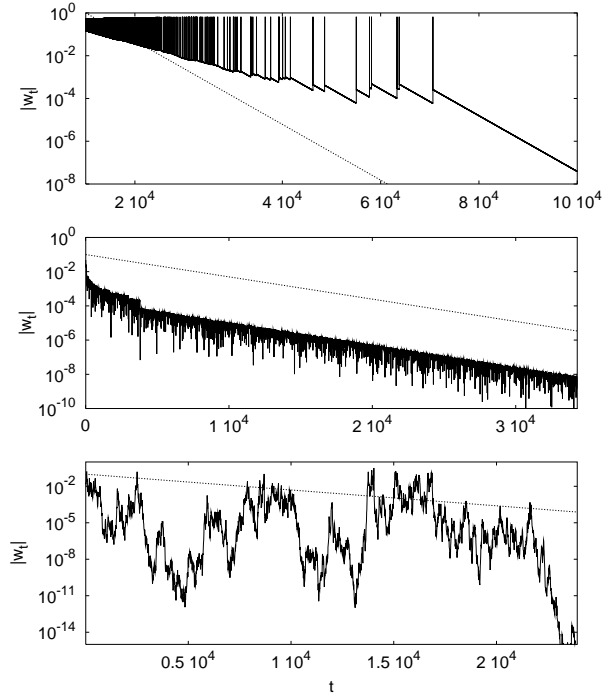} 
\caption{Typical evolution of $|w_t|=|x_t-y_t|$ starting from random
initial conditions. From top: Bernoulli map with $r=1.5$, logistic map
at the crisis and the skew tent map with $a=2/3$. For all maps, the
coupling has been chosen in order to ensure $\lambda_T=-3\,10^{-4}$. The dashed
straight lines display the $e^{\lambda_T t}$ decay.}
\label{fig:wt}
\end{figure}

Representative examples of class II maps are the generalized Bernoulli
map $f(x)=r\,x \;\; {\mbox {mod}} \;1$ that is discontinuous and its
continuous version
\begin{equation}
f(x)=\left\{\begin{array}{ll}
  a_1x &0<x<x_1\\
  m(x-x_2) &x_1<x<x_2\\
  a_1(x-x_2) & x_2<x<1
\end{array}
\right. \label{eq:bsc}
\end{equation}
with $a_1=1/x_1$, $x_{1,2}=1/a\pm \delta$, $m=-1/(2\delta)$.  As
already mentioned, the peculiarity of such maps is that they are not
only unstable with respect to infinitesimal perturbations but also to
finite ones (for (\ref{eq:bsc}) this is true for sufficiently
small $\delta$ values). 

The main differences between synchronization transitions ruled by
linear mechanisms with respect to those driven by finite amplitude
instabilities can be captured already by considering the evolution of
$|w^t|$.  The typical behavior of this quantity for $\varepsilon >
\varepsilon_l$ is reported in Fig.~\ref{fig:wt} for the Bernoulli, the
logistic and the skew tent map.  For the logistic map, apart from
(short time) fluctuations due to the multiplier variations, an average
linear decrease of $\ln|w^t|$ with slope given by $\lambda_T$ is
observed.  For the Bernoulli map, despite the TLE is negative and does
not fluctuate in time, $|w^t|$ exhibits instantaneous jumps to ${\mathcal
O} (1)$ values. These jumps, which are due to finite amplitude
instabilities, are more probable for large $|w^t|$
amplitudes. Completely analogous behaviors have been found for the map
(\ref{eq:bsc}), where the multipliers do fluctuate.

As shown in Fig.\ref{fig:wt} for the skew tent map with $a > 1/2$,
$|w_t|$ displays an intermittent evolution during the transient
preceding the synchronization, that is reflected in a diffusive motion
of $\ln|w_t|$.  As a consequence long transients are observed close to
the transition.  This phenomenon, known as {\it Modulational
Intermittency}~\cite{PikoBook} is induced by fluctuations of the
finite time TLE that do not average out in the long time limit and
as a consequence can be explained by linear analysis.

For the sake of completeness, we mention that long 
synchronization transients have  been
reported also for nonlinearly coupled expanding maps~\cite{J95}.
In this model resurgences of $|w^t|$ during the transient 
are due to the presence of an invariant chaotic repelling set.
We stress that for class II maps the transient is absolutely non chaotic
i.e. their transverse dynamics in tangent space is contracting at any time.

\subsection{Finite Size Transverse Lyapunov Exponent}

To quantitatively characterize the different {\it transverse}-space
dynamics (i.e. the evolution of $w^t$ shown in Fig.~\ref{fig:wt}) for
maps of class I and II, let us now introduce the Finite Size
Transverse Lyapunov Exponent. The FSTLE, $\Lambda_T(\Delta)$,
generalizes the concept of transverse Lyapunov exponent to finite
value of the  perturbation $|w^t| = \Delta$. 

Following Aurell et al. \cite{ABCPV96} (see also Ref.~\cite{BCFV02})
we have defined the FSTLE as follows. We introduce a set of thresholds
$\Delta_n = \Delta_0 k^n$ with $n=1,\dots,N$, since on average a
transverse expanding (resp. contracting) dynamics is expected in the
desynchronized (resp. synchronized) regime, we choose $k>1$
(e.g. $k=2$) for $\varepsilon < \varepsilon_l$ and $k<1$
(e.g. $k=1/2$) for $\varepsilon > \varepsilon_l$.  First, starting
from a random initial conditions we wait for $x^t$ to relax onto its
attractor. Then we initialize a second variable $y^0$ as
$y^0=x^0+\delta_0$ by choosing an initial perturbation $\delta_0 <
\Delta_0$ (resp. $\delta_0 > \Delta_0$) if $\varepsilon
<\varepsilon_l$ (resp. if $\varepsilon >\varepsilon_l$), and let $x^t$
and $y^t$ evolve according to Eq.~(\ref{eq:2map_1}). Care should, of
course, be taken to maintain $y^0$ within the interval of definition
of the map.  During the evolution we record the time,
$\tau(\Delta_n,k)$, needed for $|w^t|$ to pass for the first time from
one threshold $\Delta_n$ to the following one $\Delta_{n+1}$ and also
the value of $|w^t|=\tilde{w}_n$ at the moment of the passage.  When
the last threshold, $\Delta_N$, is reached the system is reinitialized
with the above described procedure and this is repeated $M$ times.
The FSTLE is thus defined as
\begin{equation}
\Lambda_T(\Delta_n) = \frac{1}{\langle\tau(\Delta_n,k)\rangle} 
\left\langle\ln\left(\frac{\tilde{w}_n}{\Delta_n}\right) \right\rangle
\, ,
\label{eq:fstle}
\end{equation}
where the average $\langle[\cdot]\rangle$ is performed over the set of
$M$ different initial conditions. In the limit $\Delta \to 0$ the
FSTLE converges to $\lambda_T$. Notice that due to its definition, the FSTLE
cannot measure at the same time expansion and contraction rates,
i.e. we have limited the analysis only to consecutive 
contractions (resp. expansions) in the
synchronized (resp. desynchronized) regime. This implies that the sign of
$\Lambda_T(\Delta)$ is always negative or positive in accordance with the
investigated situation.

\begin{figure}[ht!]
\centering
\includegraphics[draft=false, scale=.6, clip=true]{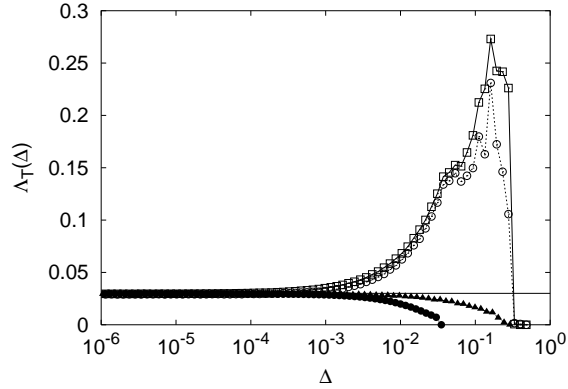} 
\caption{$\Lambda_T(\Delta)$ vs $\Delta$ in the desynchronized regime
for the Bernoulli map with $a=1.5$ (empty circles), its continuous
version (empty boxes) with $\delta=10^{-4}$, the logistic map (filled
circles) and the skew tent map (filled triangles). The coupling
constants have been chosen in such a way that in all systems the
transverse LE is equal to $\lambda_T \approx 0.03$ (solid line).}
\label{fig:fsle}
\end{figure}

In Fig.~\ref{fig:fsle} $\Lambda_T(\Delta)$ versus $\Delta$ is shown
for different maps in the desynchronized regime.  As expected, in all
cases, for very small values of $|w^t|$ the TLE is recovered. However,
at larger values of $|w^t|$ maps of the class II display an increase
of the growth rate, i.e., $\Lambda_T(\Delta)>\Lambda_T(0)$ for some
finite $\Delta$ value, while for maps of class I the FSTLE is
monotonically decreasing with $\Delta$.  $\Lambda_T(\Delta)$ provides a
quantitative measure of the strength of the nonlinear effects that, in
maps of class II, may in principle overwhelm the linear mechanisms as
pointed out in Ref.~\cite{PT94,CT01}.

\begin{figure}[ht!]
\centering
\includegraphics[draft=false, scale=.6, clip=true]{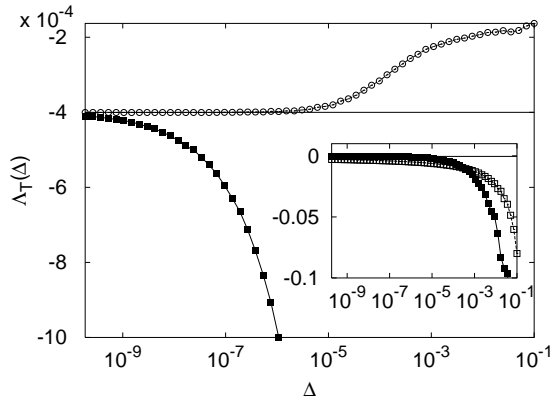} 
\caption{ $\Lambda_T(\Delta)$ vs $\Delta$ in the synchronized regime
for the Bernoulli map at $a=1.5$ (empty circles), the logistic map
(filled boxes). The inset show the logistic map (filled boxes) and the
skew tent map (empty boxes) . The straight lines correspond to the
transverse LE, here $\lambda_T= -4\,10^{-4}$ for all systems.}
\label{fig:firstarrival}
\end{figure}

In Fig.~\ref{fig:firstarrival} we report the behavior of
$\Lambda_T(\Delta)$ in the synchronized regime.  Again, while for very
small $\Delta$ the (negative) TLE is recovered, at larger $\Delta$
values the Bernoulli-like maps display an increase which is due to the
jumps ${\mathcal O}(1)$ of the transverse perturbation.
A closer inspection of the small $\Delta$ regime reveals that for the
logistic map $\Lambda_T(\Delta)\approx \lambda_T$ only for very small
scales $\sim 10^{-7}$, while for the skew tent map the asymptotic
$\Lambda_T(\Delta)\approx \lambda_T$ in never exactly reached. The
latter convergence difficulties are probably due to the intermittent behavior
along the transverse direction, in the proximity of the transition
\cite{PikoBook}. In the regime $\varepsilon <\varepsilon_l$ the finite
time effects are less important and the asymptotic behavior is 
recovered for all the investigated maps.

The results presented in this subsection clearly indicate that 
finite scale instabilities can prevail on infinitesimal ones only
for strongly nonlinear maps.
\subsection{Statistics of the synchronization times}

Notwithstanding the observed differences in the FSTLE, we found that
the critical coupling for synchronization to occur is always
given by (\ref{eq:epsc}) for both classes. This means that, at least
for the case of two coupled maps, nonlinear effects do never modify
the critical coupling.  It is natural to wonder whether  other
observables related to the synchronization transition could be
influenced by the presence of finite amplitude instabilities. As we
shall show in this section, this is the case for the synchronization
times statistics.

Let us define the synchronization time $\tau$, as the shortest time
needed to  $|w^{t}|$ for decreasing below a certain
threshold $\Theta$, with the further requirement that 
$|w^{t}| < \Theta$ for a sufficiently long successive time $T_{bt}$.
However, if the threshold value is small enough
(e.g. $\Theta \sim 10^{-8}-10^{-14}$) $\tau$ essentially
coincides with the first arrival time to the considered $\Theta$.  
The quantities of interest are the PDF's, $P(\tau)$,
and their moments (we shall focus mainly on the first moment).

\subsubsection{Class I maps}

\begin{figure}[t!]
\centering
\includegraphics[draft=false, scale=.6, clip=true, angle=0]
{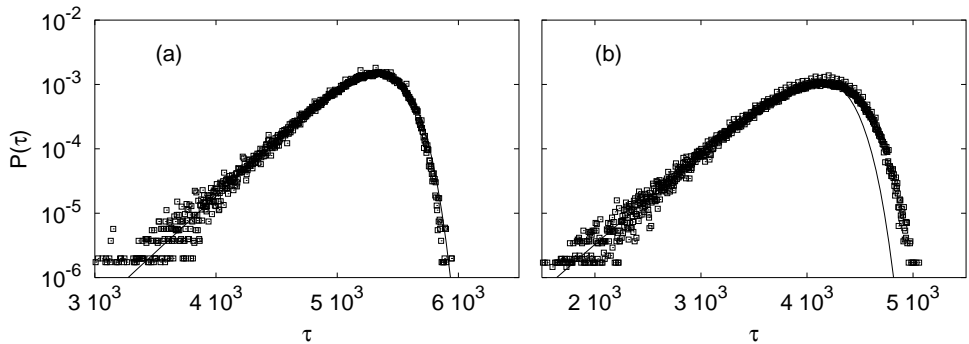} 
\caption{$P(\tau)$ as a function of the synchronization time $\tau$ :
(a) tent map for $\delta \varepsilon =(\varepsilon -\varepsilon_l) =
0.001$, the dashed line is the analytical expression
(\protect\ref{eq:p_tau}); (b) logistic map for $\delta \varepsilon =
0.001$, the solid line refers to the
expression (\protect\ref{eq:p_tau_log}) with $D=0.82$ and 
$D'=0.72$.  In all
cases the PDF have been obtained by averaging over $10^6$ different
initial conditions and by choosing $\Theta=10^{-12}$.}
\label{fig:pdftent}
\end{figure}

In Fig.~\ref{fig:pdftent} we show $P(\tau)$ for the the logistic and
the symmetric tent map.  The main feature is the presence of an
exponential tail at short times and the faster than exponential
falloff at long times.  This means that synchronization times $\tau
\gg \langle\tau \rangle$ are not observed. As a first result we
derive, by following Ref.~\cite{PG91}, an approximate analytical
expression for $P(\tau)$, which is exact for the symmetric tent map.

Let us define $z^t=\ln|w^t|$, and consider its linearized
evolution, from Eqs.~(\ref{eq:tang_diff}) and (\ref{eq:lambdat}) one
obtains
\begin{equation}
z^{t+1}=z^t + \ln|f^\prime(u^t)| +\lambda_T - \lambda_0 \, .
\label{eq:zet1}
\end{equation}
Formally the above equation can be applied only in the true linear
regime, i.e.  when $|w^t| \to 0$, so that it is not appropriate in the
early stages of the evolution. However, as shown in Fig.~\ref{fig:wt},
for maps like the logistic one, usually after a reasonably short
transient, the linear regime sets in and the use of
Eq.~(\ref{eq:zet1}) is justified for the successive evolution.  From
Eq.~(\ref{eq:zet1}) it is clear that $P(\tau)$ is related to the PDF
$P(z)$ of the variable $z$ and to that of the local multipliers
$\ln|f^\prime(u^t)|$. The formal solution of Eq.~(\ref{eq:zet1})
up to time $N$ can be written as follows:
\begin{equation}
z^N=z^0+\lambda_T N +\Lambda_N N\,, 
\label{eq:piko}
\end{equation}
where $\Lambda_N= 1/N \sum_{i=0,N-1} \ln |f^{'}(u^i)|-\lambda_0$.  For
sufficiently large $N$, large deviation theory tells us that the PDF
of $\Lambda_N$ takes the form $p(\Lambda) \sim \exp(-Ng(\Lambda))$,
being $g(\Lambda)$ the Cramer function \cite{PV87}, which is convex
and has its minimum value at $g(\Lambda\!=\!0)\!=\!0$. It is now
clear the distinction between the generic case in which $g(\Lambda)$
does not collapse onto a $\delta$-function, and the non-generic one in
which it does, as for the symmetric tent map and the logistic one at
the Ulam point.  In the former case the dynamics of $z^t$ becomes a biased
Brownian motion, with an average drift given by $\lambda_T$.

For the sake of simplicity, let us start from the tent map
(Eq.~(\ref{eq:skt}) for $a=1/2$), for which $\ln|f^\prime(u^t)| \equiv
\ln(2) \equiv \lambda_0$. In this case $\tau$ is simply given by:
\begin{equation}
\tau = \frac{z}{|\lambda_T|} - \frac{\ln \Theta}{|\lambda_T|}
\label{eq:tau}
\end{equation}
and $P(\tau)$ can be directly related to $P(z)$. Since, in the
proximity of the  transition,  $P(z=\ln|w|)$ 
assumes the form (analytically derived in Ref.~\cite{PG91}):
\begin{equation}
P(z) = \frac{2}{|\lambda_T|} {\rm e}^z {\rm
e}^{-\frac{2}{|\lambda_T|} {\rm e}^z}\,,
\label{eq:pz_tent}
\end{equation}
it can be easily obtained the related distribution
\begin{equation}
P(\tau) = 2 {\rm e}^{|\lambda_T|\tau + \ln \Theta}
{\rm e}^{-\frac{2}{|\lambda_T|}{\rm e}^{|\lambda_T| \tau + \ln \Theta}}\,,
\label{eq:p_tau}
\end{equation}
which perfectly agrees with the numerical results 
(Fig.~\ref{fig:pdftent}a).

Unfortunately the (short time) multiplier statistics of the logistic
map is nontrivial, impeding a straightforward derivation of $P(\tau)$.
However, we numerically observed that $P(w) \sim {\rm e}^{-D w}$
($w=\exp(z)$) with $D$ almost constant in a neighborhood of the
transition (namely, $D = 0.82 \pm 0.02$ for $-0.01 \le \delta
\varepsilon \le 0.01$).  This implies that
\begin{equation}
P(\tau) = 2 {\rm e}^{D|\lambda_T| \tau +  C }
{\rm e}^{\frac{2}{ D |\lambda_T|}{\rm e}^{D |\lambda_T| \tau +  C }}
\,,
\label{eq:p_tau_log}
\end{equation}
which is in fairly good agreement with the numerically evaluated PDF
(Fig.~\ref{fig:pdftent}b). Notice that $C$ is not simply given by
$D \ln\Theta$, and a fitting procedure is needed. We found
$C = D^\prime \ln \Theta$, with $D^\prime=0.72, 0.70$ and 0.65 for
$\delta \varepsilon=10^{-3},10^{-4}$ and $10^{-5}$, respectively.

\begin{figure}[t!]
\centering
\includegraphics[draft=false, scale=.64, clip=true, angle=0]
{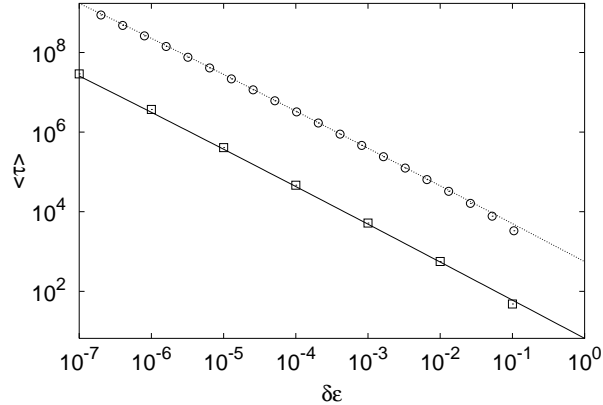} 
\caption{$\langle\tau\rangle)$ vs $\delta \varepsilon
=\varepsilon-\varepsilon_l$ for the tent map (empty boxes) and the
logistic one (empty circles). The continuous line is the prediction
(\ref{eq:tmed}) with $\Theta=10^{-12}$, the dotted one is obtained by
a best fit of the form $\langle\tau\rangle=(a+b\ln(\delta
\varepsilon))/\delta \varepsilon$, with $a=5.58$ and $b=0.24$. The
data of the logistic map have been shifted by a factor $100$ for
plotting purposes.}
\label{fig:tmedtent}
\end{figure}

The above reported results allow us also to predict the scaling of the
average synchronization time $\langle \tau \rangle$ with $\delta
\varepsilon$ in the proximity of the synchronization transition. 
In particular, for the symmetric tent map the following expression
can be derived
\begin{equation}
\langle \tau \rangle=\int_0^\infty {\rm d}\tau\, \tau P(\tau)\sim
\frac{E_1\left(\frac{2\Theta}{|\lambda_T|} \right)}{|\lambda_T|}\,,
\end{equation}
where \cite{maths}
$$
E_1(x)= \int_x^\infty {\rm d}y \, \frac{{\rm e}^{-y}}{y}=-\gamma-\ln
x-\sum_{n=1}^{\infty}{(-1)^n x^n \over n!\,n}\,,
$$
being $\gamma \sim -0.57721\dots $ the Euler-Mascheroni constant. 
Since ${2\Theta}/{|\lambda_T|}\ll 1$,  approximatively we have
\begin{equation}
\langle \tau \rangle\approx {-\ln\Theta-\gamma+\ln(|\lambda_T|/8) \over |\lambda_T|}
\, ,
\label{eq:tmed}
\end{equation}
where $|\lambda_T|\approx 2{\rm e}^{\lambda_0} \epsilon=4
\delta\varepsilon$ (being $\lambda_0=\ln 2$).  Note that
(\ref{eq:tmed}) is in perfect agreement with the numerical results
for the tent map (see Fig.~\ref{fig:tmedtent}).  For the logistic map,
a similar dependence on $\delta \varepsilon$, namely
$\langle\tau\rangle=(a+b\ln(\delta \varepsilon))/\delta \varepsilon$,
has been found.
The interesting point in (\ref{eq:tmed}) is the logarithmic correction
to the scaling $\langle \tau \rangle \sim \delta \varepsilon^{-1}$,
which would have not been predicted by a naive guess based on
(\ref{eq:tau}).

We now briefly discuss the generic situation (see Chap.~13 of
Ref.~\cite{PikoBook} for more details).  
As mentioned above,  $z^t=\ln |w^t|$ performs a biased Brownian motion,
so that computing $P(\tau)$ amounts to evaluate the distribution of 
the first passage times to a threshold of a Wiener process with drift
that, as a standard result of stochastic process theory ,is given
by an inverse Gaussian density \cite{Feller} that for long times $\tau$
becomes 
\begin{equation}
P(\tau) \sim \tau^{-3/2} e^{-\nu \tau} \quad {\mbox {with}}
\qquad \nu \propto \lambda_T^2
\,,
\label{eq:inv_gaussian}
\end{equation}
where the quadratic dependence from the TLE comes from the diffusive dynamics of
the perturbation.

\subsubsection{Class II maps}

\begin{figure}[t!]
\centering
\includegraphics[draft=false, scale=.6, clip=true]{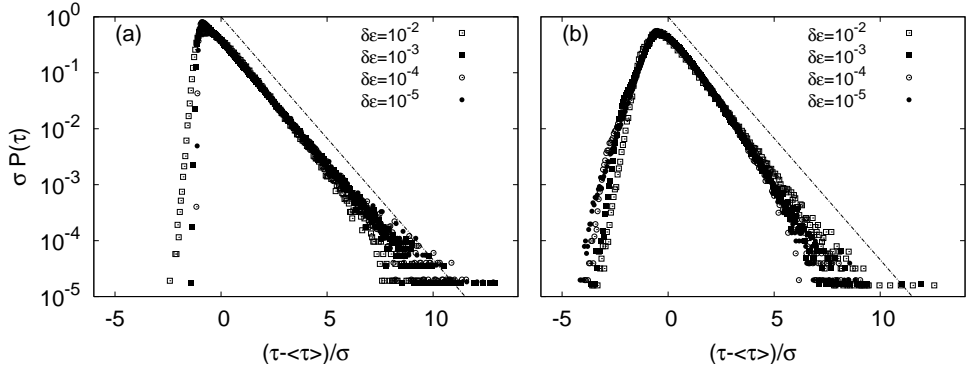}
\caption{(a)$P\sigma$ as a function of $(\tau
  -\langle\tau\rangle)/\sigma$ (where
  $\sigma=\sqrt{\langle\tau^2\rangle-\langle\tau\rangle^2}$) for the
  continuous Bernoulli shift map (\ref{eq:bsc}) with $a_1=1.1$ and
  $\delta=10^{-3}$ at different distances, $\delta\varepsilon$, from
  the critical coupling. Note the fairly good collapse, indicating a
  common decay as $e^{-x}$ (dashed line). The collapse is however only
  approximate near the peak. 
(b) Same as (a) for the
Bernoulli map with $r=2$ at
different distances from the critical coupling. The dashed line gives
$e^{-x}$.
}
\label{fig:pdfbsc}
\end{figure}

For class II maps the situation is completely different.  For both the
continuous Bernoulli map (Fig.~\ref{fig:pdfbsc}a) and the
discontinuous Bernoulli shift (Fig.~\ref{fig:pdfbsc}b) we observe that
for $\varepsilon > \varepsilon_l$ the PDF's are characterized by an
exponential tail at large $\tau$, similar to Poissonian
distributions. Moreover, we also observe that the PDF's, once rescaled
as $\sigma P(\tau)$, and reported as a function of $x= (\tau
-\langle\tau\rangle)/\sigma$ (where
$\sigma=\sqrt{\langle\tau^2\rangle-\langle\tau\rangle^2}$) collapse
onto a common curve in the proximity of the transition.  These results
are particularly striking in the case of the Bernoulli map for which
naively one would have expected a complete similarity with the tent
map, since its multiplier $f^{'}(x)=r$ is also constant.  Here the
presence of strong nonlinear effects makes the transient time
statistics to be dramatically different from (\ref{eq:p_tau}).
Indeed, as seen in Fig.~\ref{fig:wt}, $w_t$ is subject to noticeably
nonlinear amplifications induced by the (almost) discontinuities in
the map.  Therefore, the expression (\ref{eq:zet1}) is no more
appropriate to describe the dynamics of $z_t=\ln|w_t|$, and its full
nonlinear dynamics has to be taken into account. The latter is
characterized by the fact that, with a finite probability, $|w^t|$ can
jump to ${\mathcal O}(1)$ values at any time during the transient
preceding the synchronization, even if the transient is not chaotic.

This idea can be better clarified and its consequences on the $P(\tau)$
better appreciated  by considering a simple
stochastic model for the dynamics of the transverse variable at
$\varepsilon > \varepsilon_l$. This model was originally proposed in
Ref.~\cite{GLP02}, and reads as
\begin{equation}
w^{t+1}= \left\{\begin{array}{ll}
  1 &{\rm w.p.} \quad p={\rm e}^{\lambda_T}w^t \\
    {\rm e}^{\lambda_T}w^t &{\rm w.p.} \quad 1-p
\end{array}
\right.\,,
\label{eq:dsm}
\end{equation}
where $\lambda_T < 0$ and $w.p.$ is a shorthand notation for with probability.
The underlying idea is very simple: the transverse
perturbation $w^t$  is usually contracted, but with probability
proportional to its amplitude can be re-expanded to $O(1)$ values, in
agreement with the numerical observations. The jumps
occur when $x^t$ and $y^t$ are close but located at
the opposite sides of the (almost) discontinuity.

For this simple model, it is possible to derive the corresponding
probability density $P(\tau)$, which displays the same peculiar
features of the PDFs reported in Figs.~\ref{fig:pdfbsc}a 
and \ref{fig:pdfbsc}b. In particular, the long time tail.
First, notice that the
probability that an initial perturbation $w_0$ is never amplified to
${\mathcal O}(1)$ up to time $t$ is given by
\begin{eqnarray}
Q(t,w_0)\!=\!\prod_{n=1}^t [1- w_0 {\rm e}^{\lambda_T n}] \to\! \bar Q(w_0)
\!=\!\lim_{t \to
\infty} \exp\left[ -\sum_{k=1}^\infty \frac{w_0^k {\rm
e}^{-|\lambda_T| k}}{k(1- {\rm e}^{-|\lambda_T| k})}\right] .
\label{eq:p0}
\end{eqnarray}
Interestingly, for $-\infty < \lambda_T < 0$ the quantity
$\bar Q(w_0)$ is always positive and strictly smaller than $1$ for
any perturbation $|w_0| >0$, i.e. the probability that a perturbation
of amplitude $|w_0|$ can be amplified is finite at any time. 

The PDF of the times $\tau$ needed to observe
$w^\tau = \Theta$  can be factorized as
\begin{equation}
P_\Theta(\tau)=  G(\tau-\bar n) Q(\bar n,1)\,,
\label{eq:ptau_teo}
\end{equation}
where $G(\tau-\bar n)$ is the probability to receive a kick at
time $\tau-\bar n$, and $Q(\bar n,1)$ is the probability
of not being amplified for the following $\bar n$ steps,
where $\bar n$ is given by 
$$ 
\bar n = - \ln{\Theta}/|\lambda_T|\,.
$$ 
As shown in Ref.\cite{GLP02} $G(x) \sim \exp{(-\nu x)}$, and therefore
at large $\tau$'s Eq.~(\ref{eq:ptau_teo}) can be rewritten as
\begin{equation}
P_\Theta (\tau) \propto Q(\bar n,1) \exp{[-\nu(\tau+\ln{\Theta}/|\lambda_T|)]}
\label{eq:ptau_teo_2}
\end{equation}
that confirms the Poissonian character of the PDF $P(\tau)$ for the
maps of class II. By properly normalizing Eq.~(\ref{eq:ptau_teo_2}),
one obtains $\langle\tau\rangle = 1/\nu - \ln{\Theta}/|\lambda_T|$,
that is in good agreement with the numerical results for the Bernoulli
map (in particular, we considered $r=2$ and $\Theta = 10^{-12}$).
Moreover, in the interval $\delta \varepsilon \in [10^{-5};10^{-2}]$
we found that the decay rate of the PDF is directly proportional to
the TLE, i.e $\nu \approx 0.2 |\lambda_T|$, this should be contrasted
with the quadratic dependence found in the case of class I maps, see
Eq.~(\ref{eq:inv_gaussian}).

\subsection{Synchronization transition}

We conclude the investigation of two coupled maps by studying the
nature of the transition. For this purpose let us introduce the order
parameter $\Omega (\varepsilon)$ defined as (see also
Ref.~\cite{LD03}):
\begin{equation}
\Omega (\varepsilon)  =  \lim_{N\to \infty} {1\over N} \lim_{T\to \infty}
{1\over T} \sum_{i=1}^{N} \sum_{t=1}^{T} |w^{t+T_{w}}| \,,
\label{eq:op}
\end{equation}
i.e. for each value of the coupling $\varepsilon$, one considers $N$
different random initial conditions and each one is iterated for a
transient $T_{w}$ after which the time average of $|w^t|$ over a time
lapse $T$ is considered and further averaged over all the initial
conditions.  The value of the coupling giving the synchronization
transition is then implicitly defined as $\Omega (\varepsilon_{nl})=0$. In
principle, $\varepsilon_{nl}$ may differ from $\varepsilon_l$ defined
by (\ref{eq:epsc}), since this expression is valid only within a
linear approximation formalism.
\begin{figure}[ht!]
\centering
\includegraphics[draft=false, scale=.5, clip=true]{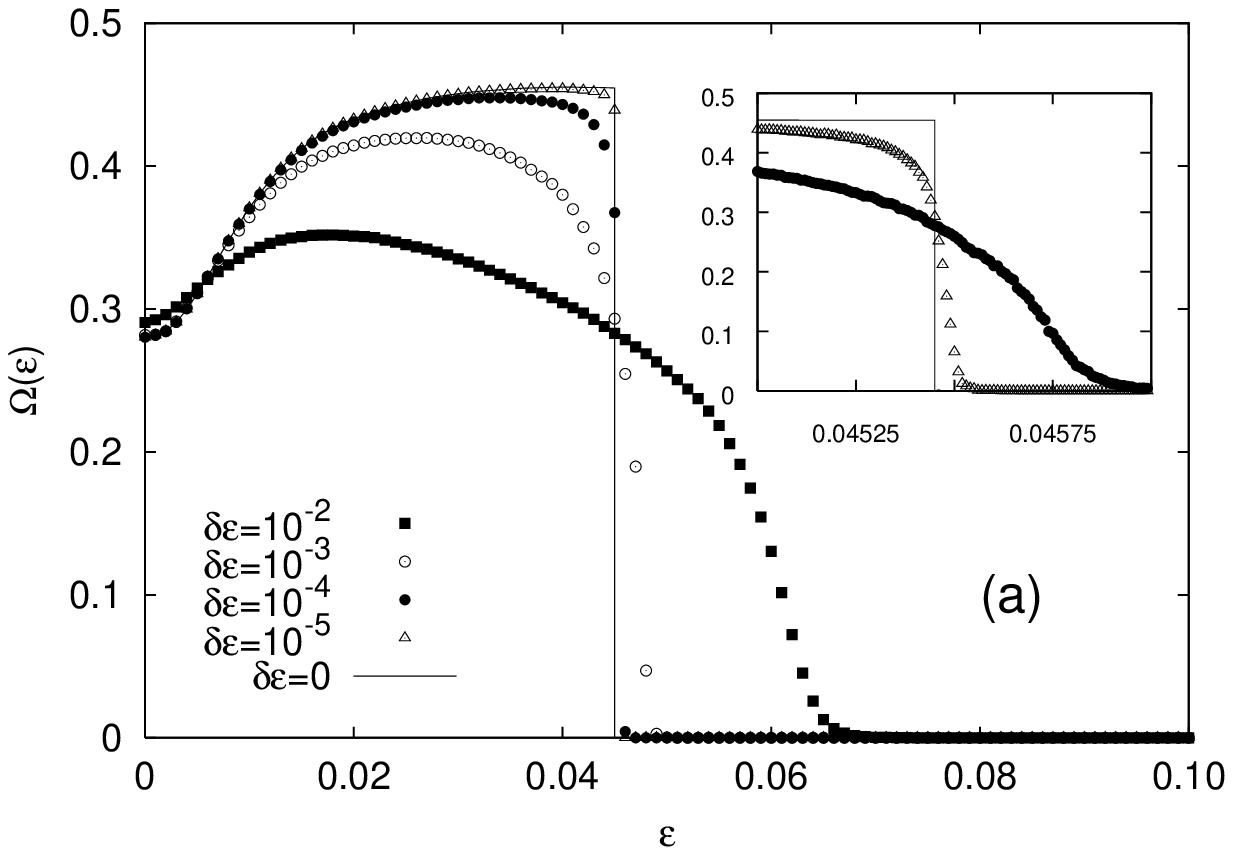} 
\includegraphics[draft=false, scale=.5, clip=true]{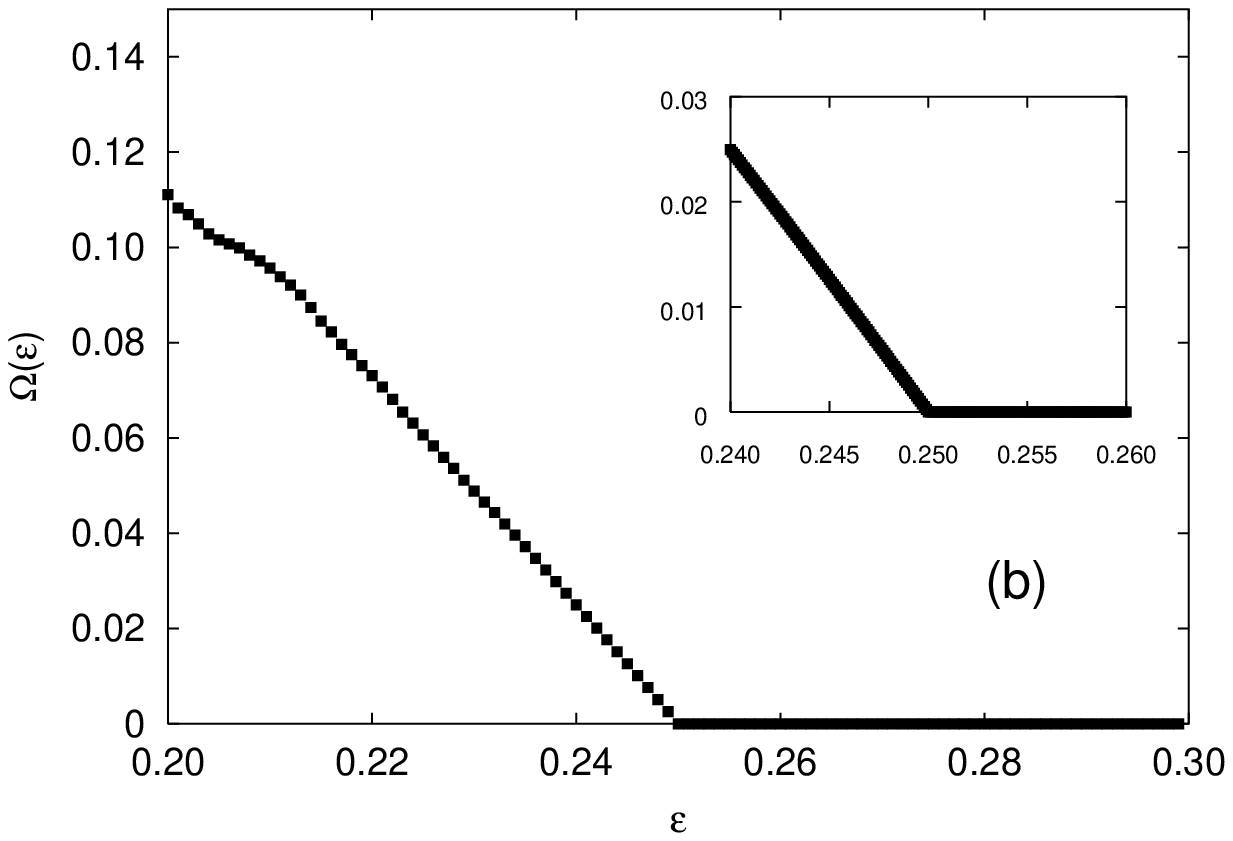} 
\caption{(a)  $\Omega(\varepsilon)$ vs $\varepsilon$ for the
Bernoulli map with $r=1.1$ and its continuous version (\ref{eq:bsc})
for different values of $\delta$.  For the Bernoulli map one has that
according to (\ref{eq:epsc}) $\varepsilon_l=0.0454545..$. The order
parameter is computed according to (\ref{eq:op}) averaging over
$N=10^3$ different random initial conditions, discarding the first
$10^6$ iteration and averaging over the following $10^5$ iterations.
$\Omega (\varepsilon)$ has been estimated with an
$\varepsilon$-resolution $10^{-3}$. The inset shows an enlargement of
the critical region done with a coupling resolution $10^{-4}$, the
same number of initial conditions and a longer integration transient
$T_w = 3\,10^6$ iterations.  (b) The same for the logistic map at the
crisis. Here (\ref{eq:epsc}) predicts $\varepsilon_l=0.25$. The blow
up of the critical region (in the inset) shows that the transition is
continuous. Note also that for $\delta \varepsilon \ll 1$ we observe
$W \propto \delta \varepsilon $.}
\label{fig:op}
\end{figure}
In Fig.~\ref{fig:op} we show the behavior of the order parameter
(\ref{eq:op}) as a function of $\varepsilon$ for both classes of
systems.  Two observations are in order. First, the transition 
to the synchronized state always occurs at the
critical coupling defined by (\ref{eq:epsc})
(i.e. $\varepsilon_{nl} \equiv \varepsilon_l$). Second, the transition is
continuous for the Logistic map (simulations show that also the tent
map, which belongs to the same class, has a continuous transition) and
discontinuous for the Bernoulli shift. This  confirms the
results reported in Ref.~\cite{LD03}. However, since the transition
for the continuous version of the Bernoulli shift map (\ref{eq:bsc}) is
steep but clearly continuous (see the inset of 
Fig.\ref{fig:op}a), this does not seem to be a generic property for all
the maps of class II.

The behavior of the Bernoulli shift map can be easily understood by 
considering the map controlling the dynamics of  $w^t$, i.e.
\begin{equation}
w^{t+1}= \left\{ \begin{array}{ll}
(1-2\varepsilon)\, r w^t    & \left\{\begin{array}{ll}
                            x^t \in I_0 & y^t\in I_0\\
                            x^t \in I_1 & y^t\in I_1
                           \end{array}\right.\\
(1-2\varepsilon)\, (r w^t +1) & x^t \in I_0\,,\quad y^t\in I_1\\
(1-2\varepsilon)\, (r w^t -1) & x^t \in I_1\,,\quad y^t\in I_0
\end{array}
\right.
\end{equation} 
being $I_0\equiv\, [0:1/r]$ and $I_1\equiv\,]1/r:1]$.  From the above
expression, it is evident that the attractor in the $(w^t,w^{t+1})$
plane has always a finite width along the transversal direction,
unless $x^t =y^t$.  This explains the discontinuity observed at
$\varepsilon_l$.  In the continuous map (\ref{eq:bsc}) the attractor
in the plane $(w^t,w^{t+1})$ has a ``transverse'' width that decreases
continuously to zero for $\varepsilon \to \varepsilon_l$.  In
conclusion, the discontinuity in the transition is a pathology of the
Bernoulli map, and possibly of other discontinuous maps.

The order parameter (\ref{eq:op}) can be seen as the time average, 
once discarded an initial transient, of the following quantity
\begin{equation}
W(t) =\langle|w^t|\rangle \qquad (\to \Omega(\epsilon)
\quad{\mbox{for}}\quad t\to \infty)\,,
\label{eq:wt}
\end{equation}
where $<\cdot>$ indicates the average over many different initial
conditions. Once $w$ is initialized ${\mathcal O}(1)$, depending on
whether $\varepsilon$ is smaller or larger than $\varepsilon_l$ two
different asymptotic behavior are observed.  In the desynchronized
regime, obviously $W(t)$ goes to a finite value $\Omega
(\varepsilon)$, while above the synchronization transition $ W(t) \to
0$ with a decay law determined by the nature of the considered maps.
For class I maps, the decay is ruled by the TLE: $W(t) \sim
\exp(\lambda_T t)$ with $\lambda_T < 0$.  For class II maps $W(t)$
shows an initial exponential decay $\sim \exp(-\mu t)$ (with $\mu>0$),
due to the effect of the resurgences averaged over many different
initial conditions, followed by a final linear decay $\exp(\lambda_T
t)$.

The nonlinear rate $\mu$ can be simply related to the exponential
tail of the PDF of the first arrival times, by assuming  
(as indeed observed) the following decay 
$P_\Theta (\tau) \propto
\exp{[-\nu(\tau -\frac{\ln{\Theta}}{|\lambda_T|})]}$.
Since $W(t)$ is the average amplitude value of the $w^t$ at time $t$,
one has 
\begin{eqnarray}
W(t) \!=\!\int_0^1\!\! d\Theta \enskip \Theta \enskip
P_\Theta(t) \propto  {\rm e}^{-\nu t}\int_0^1\!\! d\Theta \enskip
\Theta \enskip {\rm e}^{-\nu
\frac{\ln{\Theta}}{|\lambda_T|}}\!=\! 
e^{-\nu t} \int_0^1\!\! d\Theta \enskip \Theta^{1 -
\frac{\nu}{|\lambda_T|}}\,.
\label{eq:Wav}
\end{eqnarray}
The above result tells us that $\mu = \nu$ (if the integral does not 
diverge, i.e if $\nu \le |\lambda_T|$). This is confirmed by numerical
checks.

\section{Coupled maps with power law coupling}
\label{sec:3}

In this section we analyze high dimensional systems, namely the
power-law coupled maps defined in Eq.~(\ref{eq:cml}).  In particular,
for class II maps we shall show that there exist situations where, due
to the nonlinearities, the synchronization time diverges exponentially
with the number of maps.  As a consequence, the transition takes place
at a (nonlinear) critical coupling $\varepsilon_{nl}$ larger than the
linear value $\varepsilon_{l}$.  A similar phenomenon has been
observed for GCM's with nonlinear coupling, where the synchronization
time divergence is due to a chaotic transient~\cite{J95}. Chaotic
transients, diverging exponentially with the number of coupled
elements, have been reported also for spatially extended
reaction-diffusion systems~\cite{wacker} and for diluted networks of
spiking neurons~\cite{geisel04}.  However, the emphasis of our work is
on non-chaotic transients similar to the stable-chaos
phenomenon~\cite{PLOK93}.

Let us start by reviewing the linear theory developed in Ref.~\cite{anteneodo}, 
which is able to account for the synchronization transition of class I maps.

\subsection{Lyapunov Analysis}

When nonlinear effects are not sufficiently strong, excluding the
pathological cases of chaotic transients, the critical coupling for
observing the synchronization transition can be predicted by computing
the Lyapunov spectrum. In particular, since above the synchronization
transition the transverse Lyapunov exponent coincides with the second
Lyapunov exponent, $\lambda_2$, it suffices to evaluate the
dependence of the latter on $\varepsilon$.

In order to compute the Lyapunov spectrum of the model (\ref{eq:cml}),
it is necessary to consider the tangent space evolution:
\begin{eqnarray}
\delta x_i^{t+1} = (1-\varepsilon) f^\prime(x_i^t)\delta x_i^t + 
{\varepsilon \over \eta(\alpha)} \sum_{k=1}^{L^\prime}
\frac{f^\prime(x_{i-k}^t) \delta x_{i-k}^t+f^\prime(x_{i+k}^t)
\delta x_{i+k}^t}{k^\alpha}\,.
\label{eq:tang}
\end{eqnarray}
In Ref.~\cite{anteneodo} it has been shown that the full Lyapunov
spectrum can be easily obtained for the Bernoulli map:
\begin{equation}
\lambda_k= \ln r + \ln \left| 1-\epsilon +
\frac{\epsilon}{\eta(\alpha)} b_k \right|\,,
\label{eq:spectrum}
\end{equation}
where
\begin{equation}
b_k =2 \sum_{m=a}^{L^\prime} \frac{\cos(2\pi(k-1) m/L)}{m^\alpha}
\qquad k=1, \dots, L\, .
\label{eq:bk}
\end{equation}
In this notation the maximal Lyapunov exponent is given
by $\lambda_1=\ln r$ and the second Lyapunov
exponent is $\lambda_2$.

For generic maps it is still possible to obtain the Lyapunov spectrum
in an analytical way, but only in the synchronized state, by
substituting $\ln r$ with the maximal LE of the single uncoupled 
map $\lambda_0$ in (\ref{eq:spectrum}).

From a linear analysis point of view one expects that the synchronization
transition has to occur when $\lambda_2=0$. Therefore, the
following expression for the critical line in the
$(\alpha,\varepsilon)$-plane can be derived:
\begin{equation}
\epsilon_l\! =\! \left(1 \!-\! {\rm e}^{-\lambda_0} \right)
\left(1\! -\! \frac{2}{\eta(\alpha)} 
\sum_{m=1}^{L^\prime} \frac{\cos(2\pi m/L)}{m^\alpha}\right)^{-1}\,.
\label{eq:epsc_cml}
\end{equation}
As shown in Ref.~\cite{anteneodo} this prediction is well verified for the
logistic map, which is here used as a benchmark for our codes.
Let us conclude this short review by noticing that in the limit $L \to
\infty$ synchronization can be achieved only for $\alpha <
1$~\cite{anteneodo}. Therefore we shall limit our analysis to this interval.

\subsection{Nonlinear synchronization transition}

Let us now study the critical line for class II maps, exemplified by the
Bernoulli shift map and its continuous version. First of all we
introduce the main observables and the numerical method employed to
determine it.  

A meaningful order parameter for the transition is represented by time
average of the following mean field quantity :
\begin{equation}
{\rm S}(t) = \frac{1}{L} \sum_{i=1}^{L} |x_i^t-\overline{x}^t|\,,\quad
\overline{x}^t=\frac{1}{L} \sum_{i=1}^{L} x_i^t\,.
\end{equation}
This can be operatively defined as follows : firstly the system
(\ref{eq:cml}) is randomly initialized and iterated for a transient
time $T_w$ proportional to the system size $L$, then the time average
of $S(t)$ is computed over a time window $T$, i.e. $\langle {\rm S}
\rangle_T=1/T \sum_{t=1,T} {\rm S}(t)$. Finally the state of the
system is defined as synchronized if $\langle S \rangle_T < \Theta$,
being $\Theta$ a sufficiently small value ($10^{-8}-10^{-10}$ is
usually enough).  The coupling value corresponding to the
synchronization transition is then obtained by using a bisection
method: chosen two coupling values across the transition line one
corresponding to a desynchronized state ($\varepsilon_d$) and the
other to a synchronized case ($\varepsilon_s$) a third value is
selected as $\varepsilon_m=(\varepsilon_d+\varepsilon_s)/2$.  If at this new
coupling value the system synchronizes (resp. not synchronizes)
$\epsilon_m$ is identified with the new $\varepsilon_s$
(resp. $\varepsilon_d$).  The procedure is then repeated until
$(\epsilon_s-\epsilon_d)\leq \theta_\epsilon$ (in our simulations
$\theta_\epsilon=10^{-3}-10^{-5}$). Finally the critical coupling is
defined as $\epsilon_{nl}=(\epsilon_s+\epsilon_d)/2$.  The algorithm,
tested on the logistic map, was able to recover (\ref{eq:epsc_cml})
with the required accuracy.  In general one has that, once fixed
$\alpha$, $\epsilon_{nl}$ will be a function of $T_w$ for a given $L$.

\begin{figure}[th!]
\centering
\includegraphics[draft=false, scale=.65, clip=true]{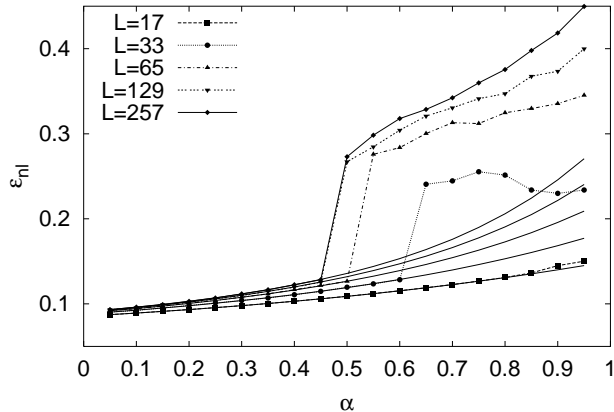}
\caption{Synchronization transition lines for the Bernoulli shift map
with $r=1.1$ at $L=17,\;33,\;65,\;129,\;257$.  The solid lines are the
analytically estimated ``linear'' transition values
Eq.~(\ref{eq:epsc_cml}), while the symbols refer to the numerically
obtained values. We used $T_w = 10^5\,L$, $T=10^3\,L$,
$\Theta=10^{-8}$ and $\theta_\epsilon=10^{-3}$.}
\label{fig:sync.bern}
\end{figure}

In Refs.~\cite{J95,anteneodo} it has been employed
as a synchronization indicator the time average of the 
following mean-field quantity:
\begin{equation}
R^t = \left| \frac{1}{L} \sum_{j=1}^L 
{\rm e}^{2 \pi i x_j^t} \right|\,.
\label{indicator}
\end{equation}
We have verified that this indicator gives results completely analogous
to $\langle S \rangle_T$ in all the considered cases.

In Ref.~\cite{anteneodo} the authors have reported for logistic
coupled maps a very good agreement between $\epsilon_l$ given by
(\ref{eq:epsc_cml}) and $\epsilon_{nl}$ estimated via $\langle R^t
\rangle_T$.  However, this is not the case for the Bernoulli map, as
shown in Fig.~\ref{fig:sync.bern}. In this case (depending on the
slope of the map $r$, on $\alpha$ and on the chain length $L$) strong
disagreements between the linear transition line given by $\epsilon_l$
and the numerically obtained values are observed.  These disagreements
are typical of class II maps.  Since the nonlinear effects locally
desynchronize the system even if $\lambda_2 <0$, in general
$\epsilon_{nl} \ge \epsilon_l$.

\begin{figure}[th]
\centering
\includegraphics[draft=false, scale=.5,clip=true]{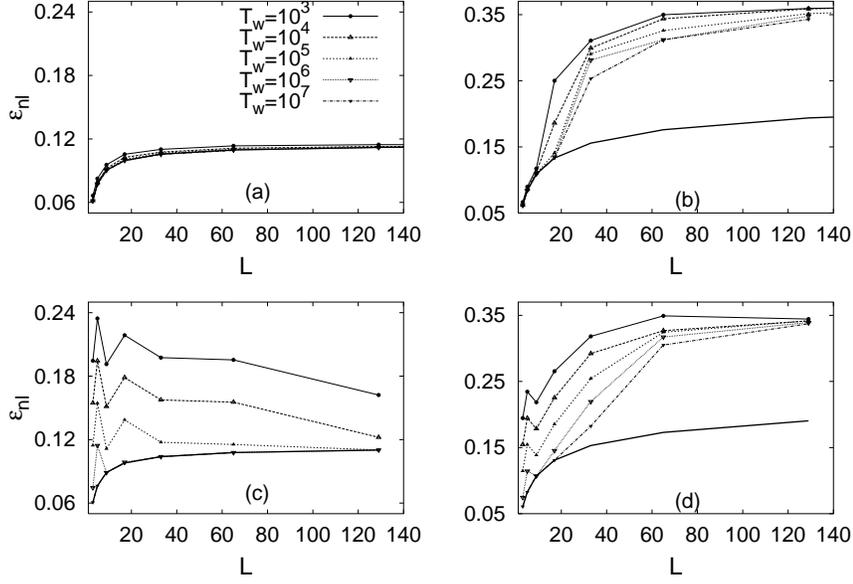} 
\caption{$\epsilon_{nl}$ as a function of $L$ for different $T_w$ in
$10^3-10^7$ for the continuous Bernoulli shift map with $r=1.1$ and
$\delta=3\times 10^{-4}$ at $\alpha=0.3$ (a) and $\alpha=0.8$ (b). The
same for the Bernoulli shift with $r=1.1$  for $\alpha=0.3$ (c) and  for
$\alpha=0.8$ (d). The continuous solid lines are the theoretical values
obtained in the linear analysis framework (\ref{eq:epsc_cml}).}
\label{fig:transients}
\end{figure}

In obtaining the data reported in Fig.~\ref{fig:sync.bern}, the cpu
time restrictions forced us to employ a large but somehow limited
transient time (namely, $T_w = 10^5\,L$). Therefore, we checked for
the dependence of the results on $T_w$ for fixed $L$. In particular,
we measured $\epsilon_{nl}$ for two $\alpha$-values only (namely,
$\alpha=0.3$ and 0.8) for several chain lengths and transient times.
This analysis has been performed for the logistic map at the crisis,
for the Bernoulli shift map with $r=1.1$ and for its continuous
version.  The results are reported in Fig.~\ref{fig:transients}.  For
the logistic map we found that, for any value of $L$ and $\alpha$, a
relatively small value of $T_w (\approx 10^4)$ was sufficient to
observe a clear convergence of $\epsilon_{nl}$ to the linear value
$\epsilon_l$. This is also the case of the Bernoulli map and its
continuous version for $\alpha=0.3$ (though as one can see in
Fig.~\ref{fig:transients}c the discontinuous map is characterized by a
slightly slower convergence than its continuous version). On the
contrary for $\alpha=0.8$ with $L \geq 21$ we were unable to observe
synchronization even for $T_w = 10^7 L$, while at $T_w \approx 10^4 L$
the logistic map had already converged to $\epsilon_l$.  Summarizing
for $r=1.1$ we found that for $\alpha \leq 0.5$ the Bernoulli shift
map and its continuous version always converge toward the linear
critical value, while for $\alpha>0.5$ at increasing $L$ the critical
coupling $\epsilon_{nl}$ becomes more and more independent of the
transient time $T_w$ (see the right of Fig.~\ref{fig:transients} for
$L=129$).  However, for a smaller value of $r$ and for large $L$ (we
investigated the case $r=1.01$ and $L=101,501$) we observed
$\varepsilon_{nl} > \varepsilon_l$ for any value of $ \alpha \in
[0;1]$.  In particular, we stress that also in the globally coupled case
(corresponding to $\alpha=0$) we have clear evidences that
$\varepsilon_{nl} > \varepsilon_{l}$.  This is due to the fact that
for $r \to 1$ nonlinear finite amplitude instabilities becomes more
and more predominant with respect to the mechanism of linear
stabilization (see~\cite{PT94,TGP95} for more details).

On the basis of the previous results, it is natural to conjecture that
in the limit $L\to \infty$ the non-linear transition is well defined,
i.e. that the limit
\begin{equation}
\epsilon_{nl}(\alpha)=\lim_{T_w\to \infty} \lim_{L\to \infty}
\epsilon_{nl}(\alpha,L,T_w)\,.
\label{eq:limit}
\end{equation}
exists and is typically larger than $\epsilon_l(\alpha)=
\lim_{L\to \infty} \epsilon_{l}(\alpha,L)$.  Note
that in the above expression the order of the two limits is crucial,
we expect that performing at fixed $L$ the limit $T_w\to \infty$ we
should always observe a convergence to $\varepsilon_l$. However, as shown in 
the following, the times to reach
synchronization may diverge exponentially fast with $L$ in the region
just above $\varepsilon_l$ making rapidly infeasible this limit.

\subsection{Synchronization Times}
As for the case of two coupled maps, we study now the synchronization
time statistics.  In particular, we consider the system above the
linear transition line $\varepsilon > \varepsilon_l$ and measure the
first passage times $\tau$ needed for $S(t)$ (\ref{indicator}) to
decrease below a given threshold $\Theta$. In this way we determine
the corresponding PDF and the related momenta.

In the high dimensional case, it is fundamental to analyze the
dependence of the synchronization time on the the system size
$L$. However, as reported in Eq.~(\ref{eq:epsc_cml}), the critical
value $\varepsilon_l$ itself depends on $L$. Therefore to perform a
meaningful comparison of systems of different sizes we considered
situations characterized by the same linear behavior in the
transverse space, i.e. having the same value of $\lambda_2$.  This
means that for each length $L$ we chose the coupling
strength according to
\begin{equation}
\varepsilon(\lambda_2)= \varepsilon_l(L) \left( 
\frac{{\rm e}^{\lambda_0} - {\rm e}^{\lambda_2}}{{\rm e}^{\lambda_0}-1} \right)
\,.
\label{eps_l2}
\end{equation}
Further tests performed at different lengths but at
a fixed distance $\Delta \varepsilon =
\varepsilon - \varepsilon_l(L)$ from the critical line give
essentially the same results.

\begin{figure}
\centering
\includegraphics[draft=false, scale=.5,clip=true]{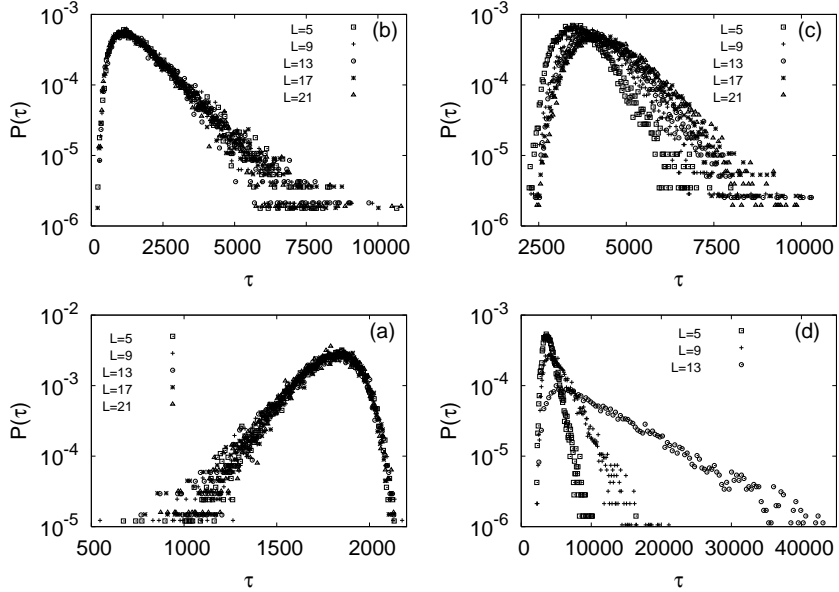}
\caption{$P(\tau)$ vs $\tau$ for different system sizes (see label) at
fixed $\lambda_2=-0.01$ for different maps: (a) logistic map at the
crisis with $\alpha=0.8$ (for $\alpha=0.3$ we obtained qualitatively
similar results); skew tent map (\protect\ref{eq:skt}) with $a=2/3$
(b); Bernoulli shift map with $r=1.1$ for $\alpha=0.3$ (c) and
$\alpha=0.8$ (d).  In the latter case sizes larger than $L=17$ were
not drawn for the sake of clarity of the plot.  The PDF's have been
obtained in all cases by considering $10^4$ different initial
conditions and with $\Theta=10^{-12}$.  }
\label{fig:pdf_log}
\end{figure}

\begin{figure}
\centering
\includegraphics[draft=false, scale=.64,clip=true]{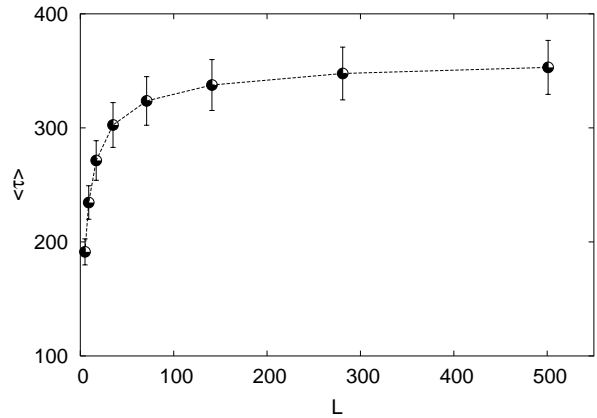}
\caption{Average synchronization time, $\langle\tau\rangle$, as a
function of the system size $L$ for the logistic map at the crisis
with $\alpha=0.6$ estimated at a fixed distance from the linear
threshold $\Delta \varepsilon=0.05$. Data have been averaged over
$10^5$ initial conditions with $\Theta=10^{-12}$.  }
\label{fig:tau_log}
\end{figure}

As shown in Fig.~\ref{fig:pdf_log}a, for the logistic map (but these
results can be extended to all the maps belonging to class I, see
Fig.~\ref{fig:pdf_log}b) the PDF's of the synchronization times
obtained at constant $\lambda_2$ display a very weak (almost absent)
dependence on the system size, and are qualitatively similar to that
found for two coupled maps (compare Fig.~\ref{fig:pdf_log}a with
Fig.~\ref{fig:pdftent}).  Moreover, measurements
of the average synchronization times $\langle\tau\rangle$, done by
fixing the ``distance'' $\Delta \varepsilon$ from the critical line,
exhibit a clear tendency to saturate for increasing $L$
(Fig.~\ref{fig:tau_log}).  Therefore, in the limit $L\to \infty$ the
synchronization time will not diverge.

For the maps of class II the situation is different.  Here, as shown
in Fig.~\ref{fig:pdf_log}c, for values of $\alpha$ sufficiently small
$P(\tau)$ is weakly dependent on the system size as found for the logistic map.
On the other hand, by considering more local couplings (e.g. $\alpha=0.8$ in
Fig.~\ref{fig:pdf_log}d), the tail of the PDF becomes more and more
pronounced as the system size increases. Results obtained by fixing the
distance from the critical coupling instead of the value of the TLE
display qualitatively similar features.

\begin{figure}[t!]
\centering
\includegraphics[draft=false, scale=.64,clip=true]{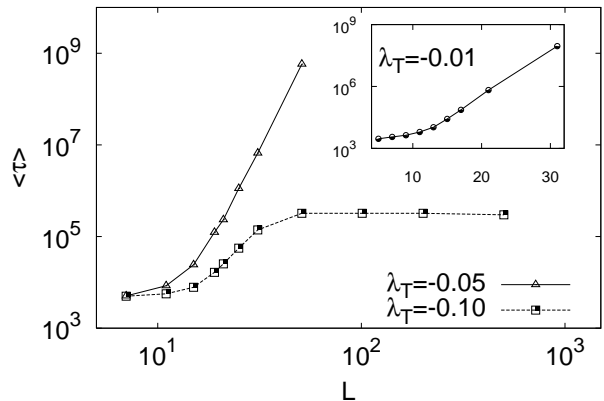}
\caption{Average synchronization times $\langle\tau\rangle$ as a
function of $L$ for the Bernoulli map with $r=1.1$ and $\alpha=0.8$
for various values of the second Lyapunov exponent (see label).  The
inset shows the case of $\lambda_2=-0.01$ in linear scale.  Data have
been averaged over $10^4$ initial conditions and $\Theta=10^{-12}$.  }
\label{fig:tau_bernou}
\end{figure}

This picture is further confirmed by examining $\langle\tau\rangle$ as
a function of $L$ for fixed $\lambda_2$ (i.e. by choosing
$\varepsilon$ according to (\ref{eps_l2})). As one can see in
Fig.~\ref{fig:tau_bernou}, $\langle\tau\rangle$ displays a dramatic
dependence on $L$.  In particular, at $\lambda_2 \ge -0.01$ (see the
inset of Fig.~\ref{fig:tau_bernou}) $\langle\tau \rangle$ grows
exponentially with $L$, while a power-like scaling is observable for
$\lambda_2 \sim -0.05$, at least for the chain lengths we could
reach. Deeper inside the ``linear'' synchronization region, i.e.  for
$\lambda_2=-0.1$, we observed a saturation of $\langle \tau\rangle$
with the system size.  It is worth stressing here that for
$\alpha=0.3$ and $r=1.1$, where no appreciable discrepancies between
$\varepsilon_{nl}$ and $\varepsilon_l$ have been observed,
$\langle\tau\rangle$ saturates for large $L$.  Nevertheless the
corresponding PDF exhibits the usual tail at long times characteristic
of maps of class II but with weak dependence on $L$ (see
Fig.~\ref{fig:pdf_log}c). Notice also that, in principle, the
transition from the exponential dependence to the saturation can be
used to estimate $\varepsilon_{nl}$ (see Ref.~\cite{baroni}), however
for the present model a systematic study of this aspect is infeasible
due to the required computational resources

We thus found clear indications that a nonlinear synchronization
transition can be observed for class II maps when finite size
nonlinear instabilities are sufficiently strong to overcome the
effects of stabilization associated with the local linear dynamics and
to the spatial coupling.  This suggests that nonlinear effects tend to
decouple the single units of the chain.  This decoupling can be indeed
related, by means of the following simple argument, to the observed
exponential divergence with $L$ of the synchronization times at
$\varepsilon_{nl} > \varepsilon > \varepsilon_{l}$.  Synchronization
happens when the difference $s_i^t=|x_i^t -\bar x^t|$ is contracted
for ever in each site until $S(t)=\bar s^t_i$ decreases below
threshold.  The probability that for a single site the initial
difference $s_i^0=w_0$ will be never amplified to ${\mathcal O}(1)$ is
given by $\bar Q(w_0)<1$ (see Eq.~(\ref{eq:p0})), therefore by
assuming that each sites is completely decoupled from its neighbors
the probability of contraction for the whole chain is given by $p=\bar
Q^L(1)=\exp{[L \ln \bar Q(1)]}$ (for simplicity we set $w_0=1$ without
loss of generality). And the probability that the system settles onto
this ``contracting'' state after $n$ steps is
\begin{equation}
P_n \simeq (1-p)^{n-1} p \sim \exp(-n p)
\label{pp_app}
\end{equation}
this quantity represents a good approximation of $P(\tau)$.  Therefore
one expects a Poissonian decay for the PDF of the synchronization
times with an associated average time $<\tau> = 1/p \propto \exp{[L
\ln(1/Q(1))]}$ exponentially diverging with the chain length, as
indeed observed.

\subsection{Properties of the transitions}

Let us now characterize the synchronization transition in the
spatially extended model (\ref{eq:cml}) within the framework of
non-equilibrium phase-transitions. A similar parallel was recently
established in a series of works concerning synchronization of two
replicas of CML's with nearest neighbor coupling
~\cite{baroni,AP02,GLP02,GLPT03}.  In these studies it has been shown
that the synchronization transition is continuous and belongs to the
Multiplicative Noise (resp. Directed Percolation) universality class
depending on the linear (resp. nonlinear) nature of the prevailing
mechanisms. In both cases one observes a transition from an active
phase (characterized by $\langle S\rangle_T =S_0 > 0$) to a unique
absorbing state (identified by $\langle S\rangle_T \equiv 0$). We
indicated with $S$ the order parameter, however for sake of clarity it
should be said that for two replicas of a CML this corresponds to
$S(t)=\sum_{i=1,L} |x^t_i-y^t_i|$ and is not given by expression
(\ref{indicator}) used in the present paper for characterizing self
synchronization. The rationale for using the same symbol is that the
two definitions embody essentially the same information: $<S>_T$ is a
measure of the density of non synchronized (active) sites.

In the proximity of the transition (if continuous)
the scaling behavior of the saturated density $S_0$ of active sites is given
by
\begin{equation}
 S_0 \sim (\varepsilon_c-\varepsilon)^{\beta}, \qquad {\rm for} \quad \varepsilon < 
 \varepsilon_c
 \enskip ,
\label{beta}
\end{equation}
while at the critical point $\varepsilon \equiv \varepsilon_c$ 
the density of active sites scales as $S(t) \sim t^{-\delta}$
and the average synchronization time diverges as
\begin{equation}
<\tau> \sim L^z
 \enskip .
\label{zeta}
\end{equation}
We denoted as $\varepsilon_c$ the critical coupling to avoid at this stage
distinctions between linearly and nonlinearly driven synchronization. 
Since continuous non-equilibrium phase transitions are typically characterized 
by three independent critical exponents, once ($\delta, \beta, z$) are known all the
other scaling exponents can be derived~\cite{haye}.

\begin{figure}
\centering
\includegraphics[draft=false, scale=.64,clip=true]{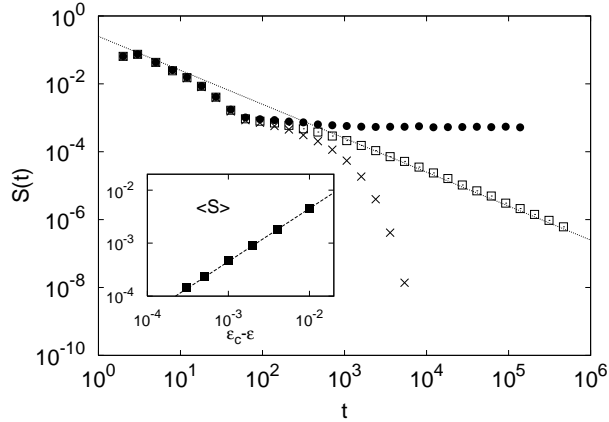}
\caption{Temporal evolution of the order parameter $S(t)$ for
$\varepsilon=\varepsilon_c-10^{-3}$ (filled circles)
$\varepsilon=\varepsilon_c$ (empty boxes) and
$\varepsilon=\varepsilon_c+10^{-3}$ (crosses). The solid straight line
displays the power law decay $t^{-1}$.  Data refer to the logistic map
with $\alpha=0.1$ and $L=501$ and are obtained after averaging over
$1,500-3,000$ initial conditions. In the inset the average value of
the order parameter $\langle S\rangle_T$ as a function of
$\varepsilon_c-\varepsilon$ is reported. The straight line shows the
scaling behavior $(\varepsilon_c-\varepsilon)^1$. Data refer to the
logistic map with $\alpha=0.3$ and $L=501$ the average is performed
over $1,000$ different initial conditions.  }
\label{fig:scaling}
\end{figure}

Let us now analyze CML's with power-law coupling (\ref{eq:cml}).  We
start with class I maps, for which the transition is completely
characterized by the linear dynamics, i.e.  $\varepsilon_c =
\varepsilon_l$. In particular, we consider the logistic maps at the
Ulam point for three different $\alpha$-values (namely, $\alpha=0.1,
0.3 $ and $0.6$). The first observation is that the critical
properties of the model seem to be independent of $\alpha$: we
measured exactly the same critical exponents for the all $\alpha$
values, namely $\delta \sim 1$, $\beta \sim 1$ (see
Fig.~\ref{fig:scaling}) and $z \sim 0$ (see Fig.~\ref{fig:tau_log}).
These exponents coincide with the mean-field exponents reported in
Ref.~\cite{HH98} for a model of anomalous Directed Percolation
(DP). The model differently from standard DP considers the spreading
of epidemics in the case of long-range infection, namely in one
spatial dimension the probability distribution for a site to be
infected at distance $r$ decays as $1/r^{1+\sigma}$.  The critical
behavior of this model can be obtained by considering a Langevin
equation with power-law decaying spatial coupling for the coarse
grained density of infected sites.  In Ref.~\cite{HH98}, the authors
found that the critical exponents vary continuously with $\sigma$, but
below a critical value ($\sigma_c=1/2$ in 1d) the exponents coincide
with the corresponding mean-field results, namely
$\beta_{MF}=\delta_{MF}=1$ and $z_{MF}=\sigma$. In the limit $\sigma
\to 0$ the latter exponents are identical to the ones we have found
for the logistic coupled maps.  In the following we shall give an
argument to explain these similarities and differences.

As shown in Refs.~\cite{baroni,AP02,GLP02,GLPT03} there is a deep
connection between the synchronization problem for two replicas of a
diffusively coupled CML and nonequilibrium phase transitions.  Indeed
there the difference field $d_i^t=|x^t_i-y^t_i|$ can be mapped onto
the density of infected (active) sites and an appropriate Langevin
equation describing the evolution of $d_i^t$ can be derived in
proximity of the transition~\cite{munoz}.  In our case of CMLs with
power-law coupling we expect that the corresponding Langevin equation
for the spatio-temporal coarse grained defect density
$s_i^t=|x^t_i-\bar x^t|$ should contain a long-range interaction with
a power law spatial coupling decaying with an exponent
$\alpha$. Therefore, in the proximity of the transition, it seems
natural to map the dynamics of (\ref{eq:cml}) onto the model studied
in ~\cite{HH98} once the exponent $\alpha$ is identified with $1-
\sigma$.  However, in (\ref{eq:cml}) the coupling is rescaled by the
factor $\eta(\alpha) \sim L^{1-\alpha}$ to avoid divergences in the
limit $L \to \infty$. The rescaling $L^{1-\alpha}$ amounts to consider
effective interaction on large scales of the type $1/r$ independently
of $\alpha$. This may explain the fact that the exponents
characterizing the transition of model (\ref{eq:cml}) coincide with
the mean-field values reported in ~\cite{HH98} for $\sigma=0$.  A
similar rescaling was performed in Ref.~\cite{TA00} to show, for a
chain of power-law coupled rotators, that all the equilibrium
properties of the system coincide for $0 \leq \alpha < 1$, once
suitably scaled.  Let us also remark that since we are dealing with a
mean-field case the nature of the noise entering in the Langevin
equation, which distinguishes Directed Percolation from Multiplicative
Noise, is irrelevant \cite{munoz_review,haye}.  This leads us to
conclude that, for continuous synchronization transitions, we would
not expect any differences in the measured exponents between linearly
and nonlinearly driven transitions.

\begin{figure}
\centering
\includegraphics[draft=false, scale=.64,clip=true]{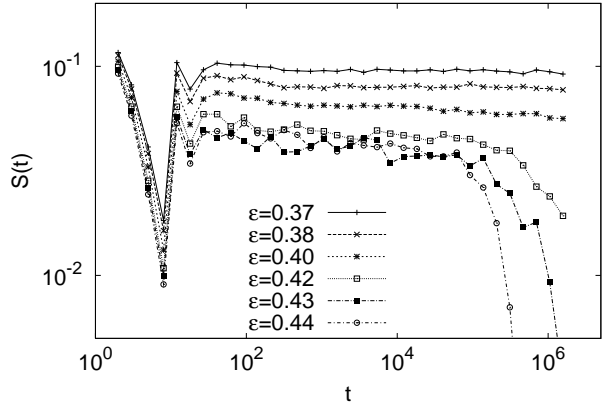}
\caption{Temporal evolution of the averaged order parameter $S(t)$ 
for the Bernoulli map with $r=1.1$, $\alpha=0.8$, $L=501$, and
different coupling strength above and below $\varepsilon_{nl}$ (see
the labels).  Average is done over $100-500$ initial conditions, tests on
shorter times and more initial conditions confirm the reliability of
the statistics. }
\label{fig:disco}
\end{figure}

In examining the critical properties associated to coupled class II
maps, we confronted with hard numerical difficulties related to the
fact that $\varepsilon_c=\varepsilon_{nl}$ cannot be measured with the
required precision. Moreover, due to the presence of exponentially
long transients in the proximity of the nonlinear transition,
simulations of model (\ref{eq:cml}) with class II maps become
extremely time consuming.  For these reasons we were unable to find
conclusive results.  However, within these limitations, our analysis
(performed up to $L=501$ and up to integration times $\sim 10^6$)
suggests that for class II maps (both the Bernoulli map and its
continuous version (\ref{eq:bsc}) with $r=1.1$ were considered) the
transition may be discontinuous as suggested from the data 
shown in Fig.~\ref{fig:disco}. In the figure
the temporal evolution of the order parameter $S(t)$
averaged over many initial conditions is reported in proximity
of $\varepsilon_{nl}$ and no indication of critical behaviour is observable.
This indicates that in class
II CMLs: short range interactions give rise to DP-like continuous
transitions, while power-law coupled maps exhibit discontinuous
transitions. We should mention that the introduction of long-range
interactions in DP model can similarly modify the nature of the
transition from continuous to discontinuous, as shown in a recent
paper \cite{haye05}.  The authors have introduced a generalized
directed percolation model, in which the activation rate of a site at
the border of an inactive island of length $\ell$ is given by
$1+a/\ell^\sigma$. In particular, in Ref.~\cite{haye05} it has been
shown that the transition is continuous for $\sigma > 1$ with exponent
coinciding with DP, and discontinuous for $0 < \sigma < 1$.

\section{Conclusions}
\label{sec:4}

In the present paper we have examined the influence of strong
nonlinearities in the transverse synchronization transition of low and
high dimensional chaotic systems.  In particular, for two coupled maps
we have shown that finite amplitude nonlinear instabilities can give
rise to long transients, preceding the synchronization, even when the
dynamics is transversally stable at any instant. This should be
contrasted with the results found for continuous maps where long
transients may happen only as a result of transient chaotic or
intermittent transverse dynamics.  In high dimensional systems with
long range (power-law) interactions strong nonlinearities may
invalidate the linear criterion to locate the critical coupling. In
this case the transition occurs, due to nonlinear mechanisms, at a
larger coupling value. The nonlinear transition is characterized by
the emergence of transients diverging exponentially with the system
size even above the linear critical coupling.  The origin of this
transition is closely related to the stable chaos phenomenon. The
synchronization phenomena both in linearly and nonlinearly driven
systems have been compared with models for long ranged contact
processes. We found that, for linearly driven systems, the transition
is continuous and the critical exponents are given by a mean field
prediction. For nonlinearly driven systems, though the results are not
conclusive, evidence of a discontinuous transition have been found.

\begin{ack}
We are grateful to W.~Just, A.~Politi and A.~Pikovsky for useful
discussions and remarks and to F.~Ginelli also for a careful
reading of this manuscript. Partial support from 
the italian FIRB contract n. RBNE01CW3M\_001 is acknowledged.
\end{ack}


\begin{thebibliography}{99}
\bibitem{epidemie} D. He and L. Stone,  
Proc R Soc Lond B Biol Sci. {\bf 270} (2003) 1519.
\bibitem{neuro} P. N. Steinmetz, A. Roy, P. J. Fitzgerald,
S. S. Hsiao, K. O. Johnson, and E. Niebur, Nature {\bf 404} (2000) 187.
\bibitem{PikoBook} A. Pikovsky, M. Rosenblum, and J. Kurths, {\it
Synchronization A Universal Concept in Nonlinear Sciences}, (Cambridge
University Press, 2001)
\bibitem{first} L.M.~Pecora and T.L.~Carroll, Phys. Rev. Lett. {\bf 64} (1990) 821.
\bibitem{KP89} S.P. Kuznetsov and A. S. Pikovsky,
Radiophys. Quantum. Electron., {\bf 32} (1989) 49.
\bibitem{PG91} A. S. Pikovsky and P. Grassberger, J. Phys. A: Math. Gen. 
{\bf 24} (1991) 4587.
\bibitem{J95} W. Just, Physica D {\bf 81} (1995) 317.
\bibitem{baroni} L.~Baroni, R.~Livi, and A.~Torcini, 
Phys. Rev. E {\bf 63} (2001) 036226.
\bibitem{AP02} V. Ahlers and A. S. Pikovsky, Phys. Rev. Lett., {\bf 88} (2002) 254101.
\bibitem{anteneodo} C. Anteneodo, S.E. de S. Pinto, A. M. Batista,
and R. L. Viana, Phys. Rev. E, {\bf 68} (2003) 045202(R);
C. Anteneodo, A.M. Batista, and R. L. Viana, Phys. Lett. A,
{\bf 326} (2004) 227.
\bibitem{neurosynch} M. Dharmala, V.K. Jirsa, and M. Ding,
Phys. Rev. Lett. {\bf 92} (2004) 028101.
\bibitem{jose}K. Wiesenfeld, P. Colet and S.H. Strogatz,
Phys. Rev. Lett. {\bf 76} (1996) 404.
\bibitem{cuore} C. Peskin, 
{\it Mathematical Aspects of Heart Physiology} (Courant Institute
of Mathematical Sciences, New York University, New York, 1975).
\bibitem{PV94}G. Paladin and A. Vulpiani, J. Phys. {\bf A 25} (1994) 4911.
\bibitem{lep-torc} A.~Torcini and S.~Lepri, Phys. Rev. E, {\bf 55} (1997) R3805.
\bibitem{gcm} K. Kaneko, Physica D, {\bf 34} (1989) 1.
\bibitem{cml} I.~Waller and R.~Kapral, Phys. Rev. A {\bf 30} , 2047
(1984); K. Kaneko, Prog. Theor. Phys. {\bf 72} (1984) 980.
\bibitem{A01} V. Ahlers, {\it Scaling and synchronization
in deterministic and stochastic nonlinear
dynamical systems}, PhD Thesis (Potsdam, 2001)
\bibitem{G03} F. Ginelli {et al.} Phys. Rev. E {\bf 68} (2003) 065102(R).
\bibitem{munoz} M. A. Mun\~oz and R. Pastor-Satorras, Phys. Rev. Lett. {\bf 90} (2003) 204101.
\bibitem{munoz_review} M. A. Mun\~oz , in
``Advances in Condensed Matter and  Statistical Mechanics",
eds. E. Korutcheva {\it et al.} (Nova Science Publishers,  2004, New York).
\bibitem{haye} H. Hinrichsen, Adv. Phys., {\bf 49}  (2000) 815.
\bibitem{ABCPV96} E. Aurell, G. Boffetta, A. Crisanti, G. Paladin and
A. Vulpiani, Phys. Rev. Lett. {\bf 77} (1996) 1262.
\bibitem{BCFV02} G. Boffetta, M. Cencini, M. Falcioni and A. Vulpiani
Phys. Rep. {\bf 356} (2002) 367.
\bibitem{CT01} M. Cencini and A. Torcini  Phys. Rev. E {\bf 63} (2001) 056201.
\bibitem{Feller} W.~Feller, {\it An introduction to probability theory
and its applications} (Wiley, New York, 1974); H.C. Tuckwell,
{\it Introduction to theoretical neurobiology - Vol. 2 -
Nonlinear and stochastic theories} 
(Cambridge University Press, Cambridge, 1988).
\bibitem{haye05} F. Ginelli, H. Hinrichsen, R. Livi, D. Mukamel,
and A. Politi, Phys. Rev. E {\bf 71} (2005) 026121.
\bibitem{PLOK93} A.~Politi, R.~Livi, G.L.~Oppo and R.~Kapral,
Europhys. Lett. {\bf 22} (1993) 571.
\bibitem{TGP95} A.~Torcini, P.~Grassberger and A.~Politi,  
J. Phys. {\bf A27} (1995) 4533.
\bibitem{PT94}A. Politi and A. Torcini, Europhys. Lett., {\bf 28} (1994) 545.
\bibitem{PV87} G.~Paladin and A.~Vulpiani, Phys. Rep. {\bf 156} (1987) 147.
\bibitem{maths} H. Jeffreys and B.S. Jeffreys,in 
{\it Methods of Mathematical Physics}, (Cambridge,
England, Cambridge University Press,1988) pp. 470-472.
\bibitem{GLP02} F. Ginelli, R. Livi, and A. Politi,
J. Phys. A:  Math. Gen. {\bf 35} (2002) 499.
\bibitem{LD03} A.~Lipowski and M.~Droz, ``Synchronization and partial
synchronization of linear maps'', cond-mat/0312067
\bibitem{wacker}  R. Wackerbauer and K. Showalter, 
Phys. Rev. Lett. {\bf 91} (2003) 174103.
\bibitem{geisel04} A. Zumdieck, M. Timme, T. Geisel, and F. Wolf,
Phys. Rev. Lett. {\bf 93} (2004) 244103.
\bibitem{GLPT03}F. Ginelli, R. Livi, A. Politi, and A. Torcini
Phys. Rev. E {\bf 67} (2003) 046217.
\bibitem{HH98} H. Hinrichsen and M. Howard,
Eur. Phys. J. B {\bf 7} (1999) 635.
\bibitem{TA00} F. Tamarit and C. Anteneodo, Phys. Rev. Lett.
{\bf 84} (2000) 208.
\end{thebibliography}
\end{document}